\documentstyle[preprint,eqsecnum,aps]{revtex}
\begin{document}
\draft
\title{Path Integral Solution by Sum Over Perturbation Series}
\author{De-Hone Lin \thanks{%
e-mail:d793314@phys.nthu.edu.tw}}
\address{Department of Physics, National Tsing Hua University \\
Hsinchu 30043, Taiwan}
\date{\today}
\maketitle
\begin{abstract}
A method for calculating the relativistic path integral solution via sum
over perturbation series is given. As an application the exact path integral
solution of the relativistic Aharonov-Bohm-Coulomb system is obtained by the
method. Different from the earlier treatment based on the space-time
transformation and infinite multiple-valued trasformation of
Kustaanheimo-Stiefel in order to perform path integral, the method developed
in this contribution involves only the explicit form of a simple Green's
function and an explicit path integral is avoided.
\end{abstract}
\newpage \tolerance=10000
\section{Introduction}
Based on the perturbation expansion of path integral formulation, Feynman
firstly introduce his famous diagram technique to give a neat interpretation
of the terms in the perturbation series and calculate the quantities of
quantum electrodynamics order by order \cite{1,2}. Over the last five
decades, Feynman's method has been successfully applied to diverse areas of
physics and achieved many accomplishments \cite{3}. Nevertheless, the exact
result of summing the perturbation series is still the aim of seeking
because of many physical effects in which non-perturbative exact result
plays the pivot's role. In this contribution, a method for calculating the
relativistic path integral is given in which the exact results only involve
the computation of some kind of moments $Q^{n}$ over the Feynman measure and
summing them in accordance with the Feynman-Kac type formula. So clear and
neat is the method that it provides us not only with an alternative approach
but a completely diverse viewpoint for treating physical problems. As an
application, we apply the formula to calculate the path integral solution of
the relativistic Aharonov-Bohm-Coulomb (A-B-C) system. It turns out that the
method presented in this paper is neat due to the avoidance of space-time
and (Kustaanheimo-Stiefel) K-S transformation in directly performming path
integral \cite{4}. The A-B-C case can serves as the prototype for the
treatment of arbitrary problems via summing the perturbation series.

\section{Path Integral Solution by Summing the Perturbation Series}

The starting point is the path integral representation for the Green's
function of a relativistic particle in external electromagnetic fields \cite
{4,5,6}: 
\begin{equation}
G({\bf x}_{b},{\bf x}_{a};E)=\frac{i\hbar }{2mc}\int_{0}^{\infty }dS\int 
{\cal D}\rho (\lambda )\Phi \left[ \rho (\lambda )\right] \int {\cal D}%
^{D}x(\lambda )\exp \left\{ -A_{E}\left[ {\bf x},{\bf \dot{x}}\right] /\hbar
\right\} \rho (0)  \label{a1}
\end{equation}
with the action 
\begin{equation}
A_{E}\left[ {\bf x},{\bf \dot{x}}\right] =\int_{\lambda _{a}}^{\lambda
_{b}}d\lambda \left[ \frac{m}{2\rho \left( \lambda \right) }{\bf \dot{x}}%
^{2}\left( \lambda \right) -i(e/c){\bf A(x)\cdot \dot{x}(}\lambda {\bf )}%
-\rho (\lambda )\frac{\left( E-V({\bf x})\right) ^{2}}{2mc^{2}}+\rho \left(
\lambda \right) \frac{mc^{2}}{2}\right] ,  \label{a2}
\end{equation}
where $S$ is defined as 
\begin{equation}
S=\int_{\lambda _{a}}^{\lambda _{b}}d\lambda \rho (\lambda ),  \label{a3}
\end{equation}
in which $\rho (\lambda )$ is an arbitrary dimensionless fluctuating scale
variable, $\rho (0)$ is the terminal point of the function $\rho (\lambda )$%
, and $\Phi \lbrack \rho (\lambda )]$ is some convenient gauge-fixing
functional \cite{4,5,6}. The only condition on $\Phi \lbrack \rho (\lambda
)] $ is that 
\begin{equation}
\int {\cal D}\rho (\lambda )\Phi \left[ \rho (\lambda )\right] =1.
\label{a4}
\end{equation}
$\hbar /mc$ is the well-known Compton wave length of a particle of mass $m$, 
${\bf A(x)}$ and $V({\bf x})$ stand for the vector and scalar potential of
the systems, respectively. $E$ is the system energy, and ${\bf {x}}$ is the
spatial part of the ($D+1$) vector $x^{\mu }=({\bf {x}},\tau )$.

The functional integral for ${\bf x}$ in representation of Eq. (\ref{a1})
can be interpreted as the expectation value of the real functional $\exp
\left\{ -\frac{1}{\hbar }\int_{\lambda _{a}}^{\lambda _{b}}d\lambda \beta
\rho (\lambda )V({\bf x(}\lambda ))\right\} $ over the measure 
\begin{equation}
K_{0}({\bf {x}}_{b},{\bf {x}}_{a};\lambda _{b}-\lambda _{a})=\int {\cal D}%
^{D}x(\lambda )e^{-\frac{1}{\hbar }\int_{\lambda _{a}}^{\lambda
_{b}}d\lambda \left[ \frac{m}{2\rho \left( \lambda \right) }{\bf {\dot{x}}}%
^{^{2}}\left( \lambda \right) -i\frac{e}{c}{\bf A}(x)\cdot {\bf \dot{x}}%
(\lambda )-\rho \left( \lambda \right) \frac{V({\bf x})^{2}}{2mc^{2}}\right]
},  \label{a5}
\end{equation}
and the entire Green's function reduces to the following formula 
\[
G({\bf {x}}_{b},{\bf {x}}_{a};E)=\frac{i\hbar }{2mc}\int_{0}^{\infty }dS\int 
{\cal D}\rho (\lambda )\Phi \left[ \rho (\lambda )\right] e^{-\frac{1}{\hbar 
}\int_{\lambda _{a}}^{\lambda _{b}}d\lambda \rho (\lambda ){\cal E}} 
\]
\begin{equation}
\times \left\langle \exp \left\{ -\frac{1}{\hbar }\int_{\lambda
_{a}}^{\lambda _{b}}d\lambda \beta \rho (\lambda )V({\bf x(}\lambda
))\right\} \right\rangle \rho (0)  \label{a6}
\end{equation}
in which ${\cal E}=$ $(m^{2}c^{4}-E^{2})/2mc^{2}$, $\beta =E/mc^{2}$ with
the notation $\left\langle \star \right\rangle $ standing for the
expectation value of the moment $\star $ over the measure $K_{0}({\bf {x}}%
_{b},{\bf {x}}_{a};\lambda _{b}-\lambda _{a})$. Eq. (\ref{a6}) forms the
basis for studying the relativistic potential problems by the Feynman-Kac
type formula. Although we have chosen the term $V({\bf x(}\lambda ))$ to
expansion, it has the aesthetic appeal on choosing convenient one according
to which the most suitable term is expanded for calculation.\ 

Expanding the potential $V({\bf x})$ in Eq. (\ref{a6}) into a power series
and interchanging the order of integration and summation, we have 
\[
G({\bf {x}}_{b},{\bf {x}}_{a};E)=\frac{i\hbar }{2mc}\int_{0}^{\infty }dS\int 
{\cal D}\rho \Phi \left[ \rho \right] e^{-\frac{1}{\hbar }\int_{\lambda
_{a}}^{\lambda _{b}}d\lambda \rho (\lambda ){\cal E}} 
\]
\begin{equation}
\times \sum_{n=0}^{\infty }\frac{(-\beta /\hbar )}{n!}^{n}\left\langle
\left( \int_{\lambda _{a}}^{\lambda _{b}}d\lambda \rho {\bf (}\lambda )V(%
{\bf x(}\lambda ))\right) ^{n}\right\rangle \rho (0).  \label{a7}
\end{equation}
We see that the calculation of path integral now turns into the computation
of the expectation value of moments $Q^{n}$ $(Q=\int_{\lambda _{a}}^{\lambda
_{b}}d\lambda \rho V({\bf x))}$ over the Feynman measure and summing them in
accordance with the Feynman-Kac type formula. Ordering the $\lambda $ as $%
\lambda _{1}<\lambda _{2}<\cdots <\lambda _{n}<\lambda _{b}$ and denoting $%
{\bf x(}\lambda _{i})={\bf x}_{i},$ the perturbation series in Eq. (\ref{a7}%
) explicitly turns into \cite{1} 
\[
\sum_{n=0}^{\infty }\frac{(-\beta /\hbar )}{n!}^{n}\left\langle \left(
\int_{\lambda _{a}}^{\lambda _{b}}d\lambda \rho {\bf (}\lambda )V({\bf x(}%
\lambda ))\right) ^{n}\right\rangle =K_{0}({\bf {x}}_{b},{\bf {x}}%
_{a};\lambda _{b}-\lambda _{a}) 
\]
\begin{equation}
+\sum_{n=1}^{\infty }\left( -\frac{\beta }{\hbar }\right) ^{n}\int_{\lambda
_{a}}^{\lambda _{b}}d\lambda _{n}\int_{\lambda _{a}}^{\lambda _{n}}d\lambda
_{n-1}\cdots \int_{\lambda _{a}}^{\lambda _{2}}d\lambda _{1}\int \left[
\prod_{j=0}^{n}K_{0}({\bf {x}}_{j+1},{\bf {x}}_{j};\lambda _{j+1}-\lambda
_{j})\right] \prod_{i=1}^{n}\rho _{i}V({\bf x}_{i})d{\bf x}_{i},  \label{a8}
\end{equation}
where $\lambda _{0}=\lambda _{a},\lambda _{n+1}=\lambda _{b},{\bf x}_{n+1}=%
{\bf x}_{b},$ and ${\bf x}_{0}={\bf x}_{a}.$

As an application of Eq. (\ref{a7}), let's apply it to the relativistic
A-B-C system in three dimensions. In this case, we have the vector and
scalar potentials 
\begin{equation}
{\bf A(x)=}2g\frac{-y\hat{e}_{x}+x\hat{e}_{y}}{x^{2}+y^{2}},\quad V(r)=-%
\frac{e^{2}}{r},  \label{a9}
\end{equation}
where $\hat{e}_{x,y}$ stands for the unit vector along the $x,y$ axis,
respectively. The perturbative expansion in Eq. (\ref{a8}) becomes 
\[
\sum_{n=0}^{\infty }\frac{(\beta e^{2}/\hbar )}{n!}^{n}\left\langle \left(
\int_{\lambda _{a}}^{\lambda _{b}}d\lambda \rho {\bf (}\lambda )\frac{1}{r}%
\right) ^{n}\right\rangle =K_{0}({\bf {x}}_{b},{\bf {x}}_{a};\lambda
_{b}-\lambda _{a})
\]
\begin{equation}
+\sum_{n=1}^{\infty }\left( \frac{\beta e^{2}}{\hbar }\right)
^{n}\int_{\lambda _{a}}^{\lambda _{b}}d\lambda _{n}\int_{\lambda
_{a}}^{\lambda _{n}}d\lambda _{n-1}\cdots \int_{\lambda _{a}}^{\lambda
_{2}}d\lambda _{1}\int \left[ \prod_{j=0}^{n}K_{0}({\bf {x}}_{j+1},{\bf {x}}%
_{j};\lambda _{j+1}-\lambda _{j})\right] \prod_{i=1}^{n}\rho _{i}\frac{d{\bf %
x}_{i}}{r_{i}}.  \label{a10}
\end{equation}
The corresponding amplitude $K_{0}$ takes the form 
\begin{equation}
K_{0}({\bf {x}}_{b},{\bf {x}}_{a};\lambda _{b}-\lambda _{a})=\int {\cal D}%
^{3}xe^{-\frac{1}{\hbar }\int_{\lambda _{a}}^{\lambda _{b}}d\lambda \left[ 
\frac{m}{2\rho \left( \lambda \right) }{\bf {\dot{x}}}^{2}\left( \lambda
\right) -i\frac{e}{c}{\bf A}(x)\cdot {\bf \dot{x}}(\lambda )-\rho \left(
\lambda \right) \frac{\hbar ^{2}}{2m}\frac{\alpha ^{2}}{r^{2}}\right] },
\label{a11}
\end{equation}
where $\alpha =e^{2}/\hbar c$ is the fine structure constant. We now choose $%
\Phi \left[ \rho \right] =\delta \left[ \rho -1\right] $ to fix the value of 
$\rho (\lambda )$ to unity. The path integral in Eq. (\ref{a7}) becomes 
\[
G({\bf {x}}_{b},{\bf {x}}_{a};E)=\frac{i\hbar }{2Mc}\int_{0}^{\infty }dSe^{-%
\frac{{\cal E}}{\hbar }S}\left\{ \vbox to 24pt{}K_{0}({\bf {x}}_{b},{\bf {x}}%
_{a};S)\right. 
\]
\begin{equation}
+\left. \sum_{n=1}^{\infty }\left( \frac{\beta e^{2}}{\hbar }\right)
^{n}\int_{\lambda _{a}}^{\lambda _{b}}d\lambda _{n}\int_{\lambda
_{a}}^{\lambda _{n}}d\lambda _{n-1}\cdots \int_{\lambda _{a}}^{\lambda
_{2}}d\lambda _{1}\int \left[ \prod_{j=0}^{n}K_{0}({\bf {x}}_{j+1},{\bf {x}}%
_{j};\lambda _{j+1}-\lambda _{j})\right] \prod_{i=1}^{n}\frac{d{\bf x}_{i}}{%
r_{i}}\vbox to 24pt{}\right\} .  \label{a12}
\end{equation}
We observe that the integration over $S$ is a Laplace transformation.
Because of the convolution property of the Laplace transformation, we obtain 
\[
G({\bf {x}}_{b},{\bf {x}}_{a};E)=\frac{i\hbar }{2mc}
\]
\begin{equation}
\times \left\{ G_{0}({\bf {x}}_{b},{\bf {x}}_{a};{\cal E})+\sum_{n=1}^{%
\infty }\left( \frac{\beta e^{2}}{\hbar }\right) ^{n}\int \left[
\prod_{j=0}^{n}G_{0}({\bf {x}}_{j+1},{\bf {x}}_{j};{\cal E})\right]
\prod_{i=1}^{n}\frac{d{\bf x}_{i}}{r_{i}}\right\}   \label{a13}
\end{equation}
with $G_{0}({\bf {x}}_{b},{\bf {x}}_{a};{\cal E})$ is the Laplace
transformation of pseudopropagator $K_{0}({\bf {x}}_{b},{\bf {x}}%
_{a};\lambda _{b}-\lambda _{a})$. Let's first analyze the influence of A-B
effect on the $G_{0}({\bf {x}}_{b},{\bf {x}}_{a};{\cal E})$. Introducing the
azimuthal angle around the A-B tube 
\begin{equation}
\varphi ({\bf x})=\arctan (y/x),  \label{a14}
\end{equation}
the components of the vector potential can be expressed as 
\begin{equation}
A_{i}=2g\partial _{i}\varphi ({\bf x}).  \label{a15}
\end{equation}
The associated magnetic field lines are confined to an infinitely thin tube
along the z-axis: 
\begin{equation}
B_{3}=2g\epsilon _{3ij}\partial _{i}\partial _{j}\varphi ({\bf x})=4\pi
g\delta ({\bf x}_{\bot }),  \label{a16}
\end{equation}
where ${\bf x}_{\bot }$ stands for the transverse vector ${\bf x}_{\bot
}=(x,y)$. Note that the derivatives in front of $\varphi ({\bf x})$ commute
everywhere, except at the origin where Stokes' theorem yields 
\begin{equation}
\int d^{2}x\left( \partial _{x}\partial _{y}-\partial _{y}\partial
_{x}\right) \varphi ({\bf x})=\oint d\varphi =2\pi .  \label{a17}
\end{equation}
The magnetic flux through the tube is defined by the integral 
\begin{equation}
\Omega =\int d^{2}xB_{3}.  \label{a18}
\end{equation}
This shows that the coupling constant $g$ is related to the magnetic flux by 
\begin{equation}
g=\frac{\Omega }{4\pi }.  \label{a19}
\end{equation}
Inserting $A_{i}=2g\partial _{i}\varphi ({\bf x})$ into the action of Eq. (%
\ref{a11}), the magnetic interaction takes the form 
\begin{equation}
A_{{\rm mag}}=i\hbar \beta _{0}\int_{0}^{S}d\lambda \dot{\varphi}(\lambda ),
\label{a20}
\end{equation}
where $\varphi (\lambda )=\varphi ({\bf x}(\lambda ))$, $\dot{\varphi}%
=d\varphi /d\lambda ,$ and $\beta _{0}$ is the dimensionless number 
\begin{equation}
\beta _{0}=-\frac{2eg}{\hbar c}.  \label{a21}
\end{equation}
The minus sign is a matter of convention. Since the particle orbits are
present at all times, their worldlines in spacetime can be considered as
being closed at infinity, and the integral 
\begin{equation}
k=\frac{1}{2\pi }\int_{0}^{S}d\lambda \dot{\varphi}(\lambda )  \label{a22}
\end{equation}
is the topological invariant with integer values of the winding number $k$.
The magnetic interaction is therefore purely topological, its value being%
\label{1} 
\begin{equation}
A_{{\rm mag}}=i\hbar \beta _{0}2k\pi .  \label{a23}
\end{equation}
The influence of A-B effect in the Green's function $G_{0}({\bf {x}}_{b},%
{\bf {x}}_{a};{\cal E})$ is as follows. In the lacking of \ A-B effect, the
Green's function 
\begin{equation}
G_{0}({\bf {x}}_{j+1},{\bf {x}}_{j};{\cal E})=\frac{m}{\hbar
(r_{j+1}r_{j})^{1/2}}\sum_{l=0}^{\infty }\sum_{k{\bf =-}%
l}^{l}g_{l}^{(0)}(r_{j+1},r_{j};{\cal E})Y_{lk}({\bf \hat{x}}%
_{j+1})Y_{lk}^{\ast }({\bf \hat{x}}_{j}),  \label{a24}
\end{equation}
where $Y_{lk}({\bf \hat{x}})$ is the 3-dimensional spherical harmonics $%
Y_{lk}({\bf \hat{x}})$ and the $g_{l}^{(0)}$ is the radial Green's function
of a particle moving in a centrifugal potential given by \cite{7,7a} 
\begin{equation}
\int_{0}^{\infty }\frac{dS}{S}e^{-\frac{{\cal E}}{\hbar }%
S}e^{-m(r_{j+1}^{2}+r_{j}^{2})/2\hbar S}I_{\sqrt{(l+1/2)^{2}-\alpha ^{2}}%
}\left( \frac{m}{\hbar }\frac{r_{j+1}r_{j}}{S}\right) .  \label{a25}
\end{equation}
The notation $I$ denotes the modified Bessel function. With the following
formulas [p.166, p.210, p212, \cite{8}], 
\begin{equation}
\left\{ 
\begin{array}{l}
P_{\nu }^{\mu }(x)=\frac{(1+x)^{\mu /2}(1-x)^{-\mu /2}}{\Gamma (1-\mu )}%
\;F\left( -\nu ,1+\nu ;1-\mu ;\frac{1-x}{2}\right)  \\ 
P_{n}^{\left( \alpha ,\beta \right) }(x)=\frac{\Gamma (1+n+\alpha )}{%
n!\Gamma (1+\alpha )}F\left( -n,\alpha +\beta +n+1;\alpha +1;\frac{1-x}{2}%
\right)  \\ 
\frac{\Gamma (1+n)}{\Gamma (1+n-l)}P_{n}^{\left( -l,l\right) }(x)=\frac{%
\Gamma (1+n+l)}{\Gamma (1+n)}\left( \frac{x-1}{2}\right) ^{l}P_{n-l}^{\left(
l,l\right) }(x)
\end{array}
\right. ,  \label{a26}
\end{equation}
where $P_{\nu }^{\mu }(x)$, $P_{n}^{\left( \alpha ,\beta \right) }(x)$ are
the associated Legendre polynomial and Jacobi function and $F$ the
hypergeometric function, it is not difficult to prove the following result 
\begin{equation}
P_{l}^{k}(\cos \theta )=(-1)^{k}\frac{\Gamma (1+k+l)}{\Gamma (1+l)}\left(
\cos \theta /2\sin \theta /2\right) ^{k}P_{l-k}^{\left( k,k\right) }(\cos
\theta ).  \label{a27}
\end{equation}
The angular part of Eq. (\ref{a24}) turns into 
\[
\sum_{k{\bf =-}l}^{l}Y_{lk}({\bf \hat{x}}_{j+1})Y_{lk}^{\ast }({\bf \hat{x}}%
_{j})=\sum_{k{\bf =-}l}^{l}\frac{2l+1}{4\pi }\frac{\Gamma \left(
1+l-k\right) }{\Gamma \left( 1+l+k\right) }P_{l}^{k}(\cos \theta
_{j+1})P_{l}^{k}(\cos \theta _{j})e^{ik\left( \varphi _{j+1-}\varphi
_{j}\right) }
\]
\[
=\sum_{k{\bf =-}l}^{l}\left[ \frac{2l+1}{4\pi }\frac{\Gamma \left(
1+l-k\right) \Gamma \left( 1+l+k\right) }{\Gamma ^{2}\left( 1+l\right) }%
\right] \left( \cos \theta _{j+1}/2\cos \theta _{j}/2\sin \theta
_{j+1}/2\sin \theta _{j}/2\right) ^{k}
\]
\begin{equation}
\times P_{l-k}^{\left( k,k\right) }(\cos \theta _{j+1})P_{l-k}^{\left(
k,k\right) }(\cos \theta _{j})e^{ik\left( \varphi _{j+1-}\varphi _{j}\right)
}.  \label{a28}
\end{equation}
To go further, let's change the variable $l$ by defining $l-k=q$ into $q$.
It is easily to fine that the Green's function of Eq. (\ref{a24}) becomes 
\[
G_{0}({\bf {x}}_{j+1},{\bf {x}}_{j};{\cal E})=\frac{m}{\hbar
(r_{j+1}r_{j})^{1/2}}\sum_{q=0}^{\infty }\sum_{k{\bf =-}\infty }^{\infty
}g_{q+k}^{(0)}(r_{j+1},r_{j};{\cal E})
\]
\[
\times \left[ \frac{2\left( q+k\right) +1}{4\pi }\frac{\Gamma \left(
1+q\right) \Gamma \left( 1+q+2k\right) }{\Gamma ^{2}\left( 1+q+k\right) }%
\right] e^{ik\left( \varphi _{j+1-}\varphi _{j}\right) }
\]
\begin{equation}
\times \left( \cos \theta _{j+1}/2\cos \theta _{j}/2\sin \theta _{j+1}/2\sin
\theta _{j}/2\right) ^{k}P_{q}^{\left( k,k\right) }(\cos \theta
_{j+1})P_{q}^{\left( k,k\right) }(\cos \theta _{j})  \label{a29}
\end{equation}
with $g_{q+k}^{(0)}$ being the radial Green's function 
\begin{equation}
\int_{0}^{\infty }\frac{dS}{S}e^{-\frac{{\cal E}}{\hbar }%
S}e^{-m(r_{j+1}^{2}+r_{j}^{2})/2\hbar S}I_{\sqrt{(q+k+1/2)^{2}-\alpha ^{2}}%
}\left( \frac{m}{\hbar }\frac{r_{j+1}r_{j}}{S}\right) .  \label{a30}
\end{equation}
Let us invoke the Poisson's summation formula [p.469 \cite{8}] 
\begin{equation}
\sum_{k=-\infty }^{\infty }f(k)=\int_{-\infty }^{\infty }dy\sum_{n=-\infty
}^{\infty }e^{2\pi nyi}f(y).  \label{a31}
\end{equation}
The entire Green's function $G_{0}({\bf {x}}_{b},{\bf {x}}_{a};{\cal E})$
containing the A-B effect becomes 
\[
G_{0}({\bf {x}}_{j+1},{\bf {x}}_{j};{\cal E})=\frac{m}{\hbar
(r_{j+1}r_{j})^{1/2}}\sum_{q=0}^{\infty }\int dz\sum_{k{\bf =-}\infty
}^{\infty }g_{q+z}^{(0)}(r_{j+1},r_{j};{\cal E})
\]
\[
\times \left[ \frac{2\left( q+z\right) +1}{4\pi }\frac{\Gamma \left(
1+q\right) \Gamma \left( 1+q+2z\right) }{\Gamma ^{2}\left( 1+q+z\right) }%
\right] e^{i(z-\beta _{0})\left( \varphi _{j+1+2k\pi -}\varphi _{j}\right) }
\]
\begin{equation}
\times \left( \cos \theta _{j+1}/2\cos \theta _{j}/2\sin \theta _{j+1}/2\sin
\theta _{j}/2\right) ^{z}P_{q}^{\left( z,z\right) }(\cos \theta
_{j+1})P_{q}^{\left( z,z\right) }(\cos \theta _{j}).  \label{a32}
\end{equation}
The sum over all $k$ in Eq. (\ref{a32}) forces $z$ to be equal to $\beta _{0}
$ modulo an arbitrary integral number leading to 
\[
G_{0}({\bf {x}}_{j+1},{\bf {x}}_{j};{\cal E})=\frac{m}{\hbar
(r_{j+1}r_{j})^{1/2}}\sum_{q=0}^{\infty }\sum_{k{\bf =-}\infty }^{\infty
}g_{q+\left| k+\beta _{0}\right| }^{(0)}(r_{j+1},r_{j};{\cal E})
\]
\[
\times \left[ \frac{2\left( q+\left| k+\beta _{0}\right| \right) +1}{4\pi }%
\frac{\Gamma \left( 1+q\right) \Gamma \left( 1+q+2\left| k+\beta _{0}\right|
\right) }{\Gamma ^{2}\left( 1+q+\left| k+\beta _{0}\right| \right) }\right]
e^{ik\left( \varphi _{j+1-}\varphi _{j}\right) }
\]
\[
\times \left( \cos \theta _{j+1}/2\cos \theta _{j}/2\sin \theta _{j+1}/2\sin
\theta _{j}/2\right) ^{\left| k+\beta _{0}\right| }
\]
\begin{equation}
\times P_{q}^{\left( \left| k+\beta _{0}\right| ,\left| k+\beta _{0}\right|
\right) }(\cos \theta _{j+1})P_{q}^{\left( \left| k+\beta _{0}\right|
,\left| k+\beta _{0}\right| \right) }(\cos \theta _{j})  \label{a33}
\end{equation}
with 
\[
g_{q+\left| k+\beta _{0}\right| }^{(0)}(r_{j+1},r_{j};{\cal E})
\]
\begin{equation}
=\int_{0}^{\infty }\frac{dS}{S}e^{-\frac{{\cal E}}{\hbar }%
S}e^{-m(r_{j+1}^{2}+r_{j}^{2})/2\hbar S}I_{\sqrt{\left[ 2(q+\left| k+\beta
_{0}\right| )+1\right] ^{2}-4\alpha ^{2}}/2}\left( \frac{m}{\hbar }\frac{%
r_{j+1}r_{j}}{S}\right) .  \label{a34}
\end{equation}
Using the orthogonality relations of Jacobi polynomials [p212, \cite{8}], 
\[
\int_{-1}^{-1}dx\left( 1-x\right) ^{\alpha }\left( 1+x\right) ^{\beta
}P_{n}^{(\alpha ,\beta )}\left( x\right) P_{m}^{(\alpha ,\beta )}\left(
x\right) 
\]
\begin{equation}
=\frac{2^{\alpha +\beta +1}}{\alpha +\beta +2n+1}\frac{\Gamma \left( \alpha
+n+1\right) \Gamma \left( \beta +n+1\right) }{n!\Gamma \left( \alpha +\beta
+n+1\right) }\delta _{m,n},  \label{a35}
\end{equation}
we perform the intermediate angular part of Eq. (\ref{a13}), it yields 
\[
G({\bf {x}}_{b},{\bf {x}}_{a};E)=\frac{i\hbar }{2mc}\sum_{q=0}^{\infty
}\sum_{k{\bf =-}\infty }^{\infty }G_{q,\left| k+\beta _{0}\right|
}(r_{b},r_{a};{\cal E})
\]
\[
\times \left[ \frac{2\left( q+\left| k+\beta _{0}\right| \right) +1}{4\pi }%
\frac{\Gamma \left( 1+q\right) \Gamma \left( 1+q+2\left| k+\beta _{0}\right|
\right) }{\Gamma ^{2}\left( 1+q+\left| k+\beta _{0}\right| \right) }\right]
e^{ik\left( \varphi _{b-}\varphi _{a}\right) }
\]
\begin{equation}
\times \left( \cos \theta _{b}/2\cos \theta _{a}/2\sin \theta _{b}/2\sin
\theta _{a}/2\right) ^{\left| k+\beta _{0}\right| }P_{q}^{\left( \left|
k+\beta _{0}\right| ,\left| k+\beta _{0}\right| \right) }(\cos \theta
_{b})P_{q}^{\left( \left| k+\beta _{0}\right| ,\left| k+\beta _{0}\right|
\right) }(\cos \theta _{a}).  \label{a36}
\end{equation}
The pure radial amplitude $G_{n,\left| k+\beta _{0}\right| }(r_{b},r_{a};%
{\cal E})$ has the form 
\begin{equation}
G_{q,\left| k+\beta _{0}\right| }(r_{b},r_{a};{\cal E})=\frac{m}{\hbar }%
\frac{1}{(r_{b}r_{a})^{1/2}}\sum_{n=0}^{\infty }\left( \frac{m\beta e^{2}}{%
\hbar ^{2}}\right) ^{n}g_{q+\left| k+\beta _{0}\right| }^{(n)}(r_{b},r_{a};%
{\cal E})  \label{a37}
\end{equation}
with $g_{q+\left| k+\beta _{0}\right| }^{(n)}$ given by 
\begin{equation}
g_{q+\left| k+\beta _{0}\right| }^{(n)}(r_{b},r_{a};{\cal E}%
)=\int_{0}^{\infty }\cdots \int_{0}^{\infty }\left[ \prod_{j=0}^{n}g_{q+%
\left| k+\beta _{0}\right| }^{(0)}(r_{j+1},r_{j};{\cal E})\right]
\prod_{i=1}^{n}dr_{i}.  \label{a38}
\end{equation}
To obtain the explicit result of $g_{q+\left| k+\beta _{0}\right| }^{(n)}$,
we note that \cite{9} 
\[
\int_{0}^{\infty }\frac{dS}{S}e^{-\frac{{\cal E}}{\hbar }%
S}e^{-m(r_{b}^{2}+r_{a}^{2})/2\hbar S}I_{\rho }\left( \frac{m}{\hbar }\frac{%
r_{b}r_{a}}{S}\right) 
\]
\begin{equation}
=2\int_{0}^{\infty }dz\frac{1}{\sinh z}e^{-\kappa (r_{b}+r_{a})\coth
z}I_{2\rho }\left( \frac{2\kappa \sqrt{r_{b}r_{a}}}{\sinh z}\right) 
\label{a39}
\end{equation}
with $\kappa =\sqrt{m^{2}c^{4}-E^{2}}/\hbar c$. With the help of the
integral formula \cite{8}, 
\begin{equation}
\int_{0}^{\infty }drre^{-r^{2}/a}I_{\nu }(\varsigma r)I_{\nu }(\xi r)=\frac{a%
}{2}e^{a(\xi ^{2}+\varsigma ^{2})/4}I_{\nu }\left( a\xi \varsigma /2\right) ,
\label{a40}
\end{equation}
we obtain the result 
\[
g_{q+\left| k+\beta _{0}\right| }^{(1)}(r_{b},r_{a};{\cal E}%
)=\int_{0}^{\infty }g_{q+\left| k+\beta _{0}\right| }^{(0)}(r_{b},r;{\cal E}%
)g_{q+\left| k+\beta _{0}\right| }^{(0)}(r,r_{a};{\cal E})dr
\]
\begin{equation}
=\frac{2^{2}}{\kappa }\int_{0}^{\infty }zh(z)dz,  \label{a41}
\end{equation}
where the function $h(z)$ is defined as 
\begin{equation}
h(z)=\frac{1}{\sinh z}e^{-\kappa (r_{b}+r_{a})\coth z}I_{\sqrt{\left[
2(q+\left| k+\beta _{0}\right| )+1\right] ^{2}-4\alpha ^{2}}}\left( \frac{%
2\kappa \sqrt{r_{b}r_{a}}}{\sinh z}\right) .  \label{a42}
\end{equation}
The expression for $g_{q+\left| k+\beta _{0}\right| }^{(n)}(r_{b},r_{a};%
{\cal E})$ can be obtained by induction with respect to $n$, and is given by 
\begin{equation}
g_{q+\left| k+\beta _{0}\right| }^{(n)}(r_{b},r_{a};{\cal E})=\frac{2^{n+1}}{%
n!}\frac{1}{\kappa ^{n}}\int_{0}^{\infty }z^{n}h(z)dz.  \label{a43}
\end{equation}
Inserting the expression in Eq. (\ref{a37}), we obtain 
\[
G_{q,\left| k+\beta _{0}\right| }(r_{b},r_{a};{\cal E})=\frac{m}{\hbar }%
\frac{2}{(r_{b}r_{a})^{1/2}}
\]
\begin{equation}
\times \int_{0}^{\infty }dze^{\left( \frac{2m\beta e^{2}}{\hbar ^{2}\kappa }%
\right) z}\frac{1}{\sinh z}e^{-\kappa (r_{b}+r_{a})\coth z}I_{\sqrt{\left[
2(q+\left| k+\beta _{0}\right| )+1\right] ^{2}-4\alpha ^{2}}}\left( \frac{%
2\kappa \sqrt{r_{b}r_{a}}}{\sinh z}\right) .  \label{a44}
\end{equation}
The integration can be done by the formula [e.g. ch. 9 \cite{7}] 
\[
\int_{0}^{\infty }dy\frac{e^{2\nu y}}{\sinh y}\exp \left[ -\frac{t}{2}\left(
\zeta _{a}+\zeta _{b}\right) \coth y\right] I_{\mu }\left( \frac{t\sqrt{%
\zeta _{b}\zeta _{a}}}{\sinh y}\right) 
\]
\begin{equation}
=\frac{\Gamma \left( \left( 1+\mu \right) /2-\nu \right) }{t\sqrt{\zeta
_{b}\zeta _{a}}\Gamma \left( \mu +1\right) }W_{\nu ,\mu /2}\left( t\zeta
_{b}\right) M_{\nu ,\mu /2}\left( t\zeta _{a}\right) ,  \label{a45}
\end{equation}
where $M_{\mu ,\nu }$ and $W_{\mu ,\nu }$ are the Whittaker functions and
the range of validity is given by 
\[
\begin{array}{l}
\zeta _{b}>\zeta _{a}>0, \\ 
{Re}[(1+\mu )/2-\nu ]>0, \\ 
{Re}(t)>0,\mid \arg t\mid <\pi .
\end{array}
\]
We complete the integration and obtain 
\[
G_{q,\left| k+\beta _{0}\right| }(r_{b},r_{a};{\cal E})=\frac{1}{(r_{b}r_{a})%
}\frac{mc}{\sqrt{m^{2}c^{4}-E^{2}}}\qquad \qquad \qquad 
\]
\[
\times \frac{\Gamma \left( 1/2+\sqrt{\left[ 2(q+\left| k+\beta _{0}\right|
)+1\right] ^{2}-4\alpha ^{2}}/2-E\alpha /\sqrt{m^{2}c^{4}-E^{2}}\right) }{%
\Gamma \left( 1+\sqrt{\left[ 2(q+\left| k+\beta _{0}\right| )+1\right]
^{2}-4\alpha ^{2}}\right) }
\]
\[
\times W_{E\alpha /\sqrt{m^{2}c^{4}-E^{2}},\sqrt{\left[ 2(q+\left| k+\beta
_{0}\right| )+1\right] ^{2}-4\alpha ^{2}}/2}\left( \frac{2}{\hbar c}\sqrt{%
m^{2}c^{4}-E^{2}}r_{b}\right) 
\]
\begin{equation}
\times M_{E\alpha /\sqrt{m^{2}c^{4}-E^{2}},\sqrt{\left[ 2(q+\left| k+\beta
_{0}\right| )+1\right] ^{2}-4\alpha ^{2}}/2}\left( \frac{2}{\hbar c}\sqrt{%
m^{2}c^{4}-E^{2}}r_{a}\right) .  \label{a46}
\end{equation}
The entire solution of path integral becomes 
\[
G({\bf {x}}_{b},{\bf {x}}_{a};E)=\frac{i\hbar }{2mc}\frac{mc}{4\pi r_{b}r_{a}%
\sqrt{m^{2}c^{4}-E^{2}}}
\]
\[
\times \sum_{q=0}^{\infty }\sum_{k{\bf =-}\infty }^{\infty }\left\{ \frac{%
\Gamma \left( 1/2+\sqrt{\left[ 2(q+\left| k+\beta _{0}\right| )+1\right]
^{2}-4\alpha ^{2}}/2-E\alpha /\sqrt{m^{2}c^{4}-E^{2}}\right) }{\Gamma \left(
1+\sqrt{\left[ 2(q+\left| k+\beta _{0}\right| )+1\right] ^{2}-4\alpha ^{2}}%
\right) }\right\} 
\]
\[
\times W_{E\alpha /\sqrt{m^{2}c^{4}-E^{2}},\sqrt{\left[ 2(q+\left| k+\beta
_{0}\right| )+1\right] ^{2}-4\alpha ^{2}}/2}\left( \frac{2}{\hbar c}\sqrt{%
m^{2}c^{4}-E^{2}}r_{b}\right) 
\]
\[
\times M_{E\alpha /\sqrt{m^{2}c^{4}-E^{2}},\sqrt{\left[ 2(q+\left| k+\beta
_{0}\right| )+1\right] ^{2}-4\alpha ^{2}}/2}\left( \frac{2}{\hbar c}\sqrt{%
m^{2}c^{4}-E^{2}}r_{a}\right) .
\]
\[
\times \left\{ \frac{\Gamma \left( 1+q\right) \Gamma \left( 1+q+2\left|
k+\beta _{0}\right| \right) \left[ 2\left( q+\left| k+\beta _{0}\right|
\right) +1\right] }{\Gamma ^{2}\left( 1+q+\left| k+\beta _{0}\right| \right) 
}\right\} 
\]
\[
\times e^{ik\left( \varphi _{b-}\varphi _{a}\right) }\left( \cos \theta
_{b}/2\cos \theta _{a}/2\sin \theta _{b}/2\sin \theta _{a}/2\right) ^{\left|
k+\beta _{0}\right| }
\]
\begin{equation}
\times P_{q}^{\left( \left| k+\beta _{0}\right| ,\left| k+\beta _{0}\right|
\right) }(\cos \theta _{b})P_{q}^{\left( \left| k+\beta _{0}\right| ,\left|
k+\beta _{0}\right| \right) }(\cos \theta _{a}).  \label{a47}
\end{equation}
This result is given in Refs. \cite{4}, and p.304 \cite{7a} in the first
time where the same result must invoke the complicate space-time and
multi-valued K-S transformations to perform the path integral. In present
paper, this procedures is avoided and can be applied to arbitrary potential
problems.

\section{Concluding Remarks}

In the paper, a method for calculating the relativistic path integral
involved essentially the computation of the expectation value of convenient
moments $Q^{n}$, such as $Q=\int_{\lambda _{a}}^{\lambda _{b}}d\lambda \rho
(\lambda )V({\bf x)}$ if we expands the term in action potential term, over
the Feynman measure and summing them in accordance with the Feynman-Kac type
formula is given. As an realization, the path integral solution of
relativistic A-B-C system is given. Different from the former treatment in
Ref. \cite{4}, where the same problem must invoke the complicated space-time
and the multi-valued K-S{\it \ }transformations to perform path integral,
the merits of the method used in the paper is that it involves only the
explicit form of some known Green's function and explicit path integral is
avoided. The A-B-C system can serves as the prototype for the treatment of
arbitrary problems via summing the perturbation series. It is our hope that
our studies would help to achieve the ultimate goal of obtaining a
comprehensive and complete solutions in perturbation series based on the
path integral of quantum mechanics and quantum field theory, including
quantum gravity and cosmology.

\end{document}